\documentstyle{article}
\begin{document}
\LARGE
\begin{center}
\bf Quantum Creation of a Black Hole
\vspace*{0.7in}
\normalsize \large \rm 

Wu Zhong Chao

International Center for Relativistic Astrophysics,

Rome University, Rome, Italy

and 

Specola Vaticana, Vatican City State

\vspace*{0.2in}
\large
\bf
Abstract
\end{center}
\vspace*{.1in}
\rm
\normalsize
\vspace*{0.1in}

Using the Hartle-Hawking no-boundary proposal for the wave
function of the universe, we can study the wave function and
probability of a single black hole created at the birth of the
universe.  The black hole originates from a constrained
gravitational instanton with conical singularities. The wave
function and
probability of a universe with a black hole are calculated at the
$WKB$ level. The probability of a black hole creation is the
exponential of one quarter of the sum of areas of the black hole
and cosmological horizons. One quarter of this sum is
the total entropy of the universe. We show that 
these arguments apply to all kinds of black holes in the de
Sitter space background.

\vspace*{0.3in}

PACS number(s): 98.80.Hw, 98.80.Bp, 04.60.Kz, 04.70.Dy

e-mail: wu@axp3g9.icra.it

\vspace*{0.5in}

\pagebreak

\large \bf I. Introduction
\vspace*{0.15in}

\rm 

\normalsize

Hawking's theory of No-Boundary Universe has for the
first time led to a self-contained cosmology. Now in principle,
one can predict everything in the universe solely from physical
laws and the long-standing first cause problem has been
dispelled. In quantum
cosmology one of the most challenging problems is the existence
of primordial black holes. 

It is well known that a black hole can be formed in two ways, the
first being the gravitational collapse of a massive star. If the
mass
of a star exceeds about twice that of the Sun, a black hole will
be its ultimate corpse. The second way originates from the
fluctuation of matter distribution in the early universe. In the
big bang model, the matter content can be classically
described [1][2], while in the inflationary universe the matter
content is attributed to the quantum fluctuation of the Higgs
scalar [3].  

The discovery of Hawking radiation of black holes has been the
most
important event of gravitational physics for several decades.
However, the life of the first kind of black hole is very much
longer than the age of the universe. The only hope of confirming
Hawking radiation by observation is through primordial
black hole hunting.

Strictly speaking, black holes formed through either way
mentioned above can hardly be regarded as primordial.
A true primordial black hole should be created at the moment of
the birth of the universe.  Over the last decade there have been
several attempts to deal with this problem, however their results
are not conclusive [4][5].  

It is believed that the very early universe is approximately
described by a de Sitter metric. In quantum
cosmology, at the Planckian era, the universe was created from a
$S^4$ space through a quantum transition. Therefore, to study the
problem of primordial black hole creation in the de Sitter
spacetime background is of twofold interest, for cosmology and
for black hole physics.

There have been many studies recently on quantum creation of
charged or neutral
black hole pairs in the de Sitter spacetime background
[6][7][8][9][10][11].
The case of a single primordial black hole is the topic
of this paper. Sect. II is devoted to the theory of constrained
gravitational instanton. Sect. III considers both the neutral and
nonrotating black hole, i.e.  the Schwarzschild black hole and
the charged but nonrotating black hole, i.e.
the Reissner-Nordstr$\rm\ddot{o}$m black hole. Sect. IV
is devoted to the
rotating but neutral black hole case, i.e. the Kerr black
hole. Sect. V investigates the rotating and charged case, i.e.
the Newman black hole. By the no-hair theorem, all these kinds of
black holes have exhausted the stationary vacuum or electrovac
cases. Therefore, the problem of quantum creation of a single
black
hole in quantum cosmology is completely resolved. Sect. VI is a
discussion.

\vspace*{0.3in}

\large \bf II. The constrained gravitational instantons
\vspace*{0.15in}

\rm
\normalsize

In the No-Boundary Universe the wave function of the
universe is given by [12]
\begin{equation}
\Psi(h_{ij}, \phi) = \int_{C}d[g_{\mu \nu}] d [\phi] \exp (-
\bar{I} ([g_{\mu \nu}, \phi]),
\end{equation}
where the path integral is over class $C$ of compact
Euclidean $4$-metrics and matter field configurations, which
agree with the given $3$-metrics $h_{ij}$ of the only boundary
and matter configuration $\phi$ on it. Here $\bar{I}$ means the
Euclidean action.

The Euclidean action for the gravitational part for a smooth
spacetime
manifold $M$ with boundary $\partial M$ is
\begin{equation}
\bar{I} = - \frac{1}{16\pi} \int_M d^4x g^{1/2}(R - 2\Lambda) -
\frac{1}{8\pi} \int_{\partial M} d^3x h^{1/2} K,
\end{equation}
where $\Lambda$ is the cosmological constant, $R$ is the scalar
curvature, $K$ is the trace of the second fundamental form of the
boundary, $g$ and $h$ are the determinants of $g_{\mu \nu}$ and
$h_{ij}$ respectively.

The dominant contribution to the path integral comes from some
classical solutions of the field equations, which are the saddle
points of the path integral. 

The probability of the Lorentzian trajectory emanating from the
3-surface $\Sigma$ with the matter field $\phi$ on it can be
written as
\begin{equation}
P = \Psi^\star \Psi = \int_C d [g_{\mu \nu}] d[\phi] \exp(-
\bar{I} ([g_{\mu \nu}, \phi ]),
\end{equation}
where class $C$ is all no-boundary compact Euclidean 4-metrics
and
matter field
configurations which agree with the given 3-metric $h_{ij}$ and
matter field $\phi$ on $\Sigma$.

Here, we do not restrict class $C$ to contain regular metrics
only, since the derivation from Eq. (1) to Eq. (3) has already
led to some jump discontinuities in the extrinsic curvature at
$\Sigma$.

The main contribution to the path integral in Eq. (3) is due to
the stationary action 4-metric, which meets all requirements on
the
3-surface $\Sigma$ and other restrictions. At the $WKB$ level,
the exponential of the negative of the stationary action is the
probability of the corresponding Lorentzian trajectory.

From the above viewpoint, an extension of class $C$ to include
metrics with some mild singularities is essential. Indeed, in
some sense, the set of all regular metrics is not complete, since
for many cases, under the usual regularity conditions and the
requirements at the equator $\Sigma$, there may not exist any
stationary action metric, i.e. a gravitational instanton. It is
not
clear, how large class $C$ should be. A necessary condition for a
metric to be a member it that its scalar curvature should be
well-defined mathematically. It is reasonable to include jump
discontinuities of extrinsic curvature and their degenerate
cases, that is the conical or pancake singularities. For this
kind of
singularity, the quantity $g^{1/2}R$ can be interpreted as a
distribution-valued density [13].

Although the regularity conditions on the 4-metrics and the
requirements from the equator $\Sigma$ sometimes are so strong
that no gravitational instanton exists, one can still hopefully
find a stationary action nonregular solution with some mild
singularities within class $C$, which can be called the
constrained gravitational instanton. Here the manifold of the
instanton is constrained by the equator $\Sigma$.

It has been proven [13] a stationary action regular solution
keeps
its status
under the extension of class $C$. However, if a stationary action
regular
solution cannot be found, then it can probably be expected with
some
singularities in class $C$.
For a model with $S^1 \times S^2$ topology under the
minisuperspace ansatz
\begin{equation}
ds^2 = a^2(r) dr^2 + b^2(r) d \tau^2 + c^2(r) d \Omega^2_2,
\end{equation}
where $z$ is periodic with period $2\pi$, and $d\Omega^2_2$
represents the metric of a unit 2-sphere, the solution satisfies
the usual
Einstein field equation except for the singularities at the final
$r = r_f$ and initial $r = r_i$ surfaces. One can rephrase this
by saying that the solution obeys the generalized Einstein
equation in the whole manifold. Since this result is derived from
first principles, one should not feel upset about this situation.

Except for the interpretation of probability the above arguments
can also be applied to the Lorentzian regime with a purely
imaginary phase. The dominating contribution is again due to the
stationary action trajectories. However, in most cases, the
restrictions are not too strong, and one can find a regular
metric satisfying the usual Einstein field equation.

At the transition surface $\Sigma$, it is assumed that along
neither of 
the sides of
$\Sigma$ does a singular matter distribution exist. It follows
from the
Einstein equation that the fundamental form $K_{ij}$ at $\Sigma$
should vanish,
\begin{equation}
K_{ij} = 0.
\end{equation}
This condition cannot apply to the mild singularities at $\Sigma$
if there is
any, since the usual Einstein equation does not hold there.
 
The singularity problem associated with a gravitational instanton
is not always disturbing; in fact it can be beneficial. If the
restrictions are weak enough to allow a regular instanton, then
the Lorentzian evolution originating from it must be most
probable one. Therefore, in order to find the most probable
Lorentzian evolution, one needs only find a regular instanton,
which then identifies the 3-metric and matter field on $\Sigma$.

In general, the wave packet of a wave function of the universe
represents an ensemble of classical trajectories. Under our
scheme, the most probable trajectory associated with an instanton
can be singled out [14]. Thus, quantum cosmology obtains its
complete power of prediction. It means there is no more degree of
freedom left as long as the model is well-defined.

On the other hand, the more severe the restrictions are, the
larger the
stationary action is, and therefore,  the less probable its
corresponding
Lorentzian evolution. This is the situation with a constrained
instanton. We shall see this in the case of a primordial black
hole.

If there is no black hole in the universe, then one can get a
regular instanton $S^4$. If there is, then the restrictions are
strong enough to forbid regular solution. Therefore, the
probability of a universe without a black hole is always greater
than one with a black hole. Our calculation will
support this.

There has been some progress in this direction. However, nearly
all scenarios studied are associated with pair creation of black
holes [6][7][8][9][10][11]. The main reason for this is that,
people
consider our universe to have been
created by a quantum transition from a gravitational instanton.
There does not exist any gravitational instanton
which provides the seed for the creation of a single black hole
in the de Sitter background.

As we mentioned above, in quantum cosmology one uses a
Lorentzian metric to join a Euclidean metric, both being sectors
of a complex manifold. However, there exist very few complex
manifolds satisfying the Einstein equation with both a Euclidean
and a Lorentzian sectors [15]. One may appeal to some
approximately Euclidean or Lorentzian sectors, but only at the
price of losing some of the beauty of the theory. In the extended
framework the
requirement becomes quite loose. The situation of black hole
creation we are going to investigate is the best illustration.
\vspace*{0.3in} 
 
\large \bf III. The spherically symmetric black hole 
\vspace*{0.15in} 
 
\rm  
\normalsize

Let us begin with a quantum spherically symmetric vacuum or
electrovac model 
with a positive cosmological constant $\Lambda$. The cosmological

constant may be effective due to the Planckian inflation in the 
Hawking massive scalar model [16]. At the semiclassical level
the 
evolution of the universe is described by its classical 
solutions.  The Schwarzschild-de Sitter spacetime with mass 
parameter $m$ and zero charge $Q$ is the unique spherically
symmetric vacuum solution 
to the Einstein equation with a cosmological constant $\Lambda$. 
The Reissner-Nordstr$\rm\ddot{o}$m-de Sitter spacetime, with mass
parameter $m$, nonzero charge $Q$ and a cosmological constant
$\Lambda$, is the only spherically
symmetric electrovac solution to the Einstein and Maxwell
equations. 
Its Euclidean metric can be written as [17] 
\begin{equation} 
ds^2 = \left (1- \frac{2m}{r} +\frac{Q^2}{r^2} - \frac{\Lambda
r^2}{3} \right)d\tau^2  
+ \left (1- \frac{2m}{r} + \frac{Q^2}{r^2}- \frac{\Lambda
r^2}{3}\right )^{-1}dr^2 
+ r^2 (d\theta^2 + \sin^2 \theta d\phi^2).
\end{equation}

We can set 
\begin{equation} 
V_s =  1- \frac{2m}{r} +\frac{Q^2}{r^2}- \frac{\Lambda r^2}{3}. 
\end{equation} 
 
For convenience one can make a factorization  
\begin{equation} 
V_s = - \frac{\Lambda}{3r^2} (r - r_0)(r - r_1)(r - r_2)(r -
r_3), 
\end{equation} 
where $r_0, r_1, r_2, r_3$ are in ascending order. $r_2$ and
$r_3$ are 
the black hole and cosmological horizons, where conical
singularities may occur, $r_0$ is negative. If
the black hole is neutral, then $r_1$ can be set to zero, and
there are essentially three roots left.

The gauge field is
\begin{equation}
 F= - \frac{iQ}{r^2} d\tau \wedge dr
\end{equation}
for an electrically charged solution, and
\begin{equation}
F= Q\sin \theta d \theta \wedge d \phi
\end{equation}
for a magnetically charged solution. We shall not consider dyonic
solutions. 

The roots satisfy the following relations
\begin{equation}
\sum_i r_i = 0,
\end{equation}
\begin{equation}
\sum_{i >j} r_i r_j = -\frac{3}{\Lambda},
\end{equation}
\begin{equation}
\sum_{i>j>k} r_i r_j r_k = -\frac{6m}{\Lambda}
\end{equation}
and
\begin{equation}
\prod_i r_i = - \frac{3Q^2}{\Lambda}.
\end{equation}
 
The black hole and cosmological surface gravities $\kappa_2$ and 
$\kappa_3$ are [13] 
\begin{equation} 
\kappa_2 = \frac{1}{2}|V_{s}^\prime(r_2)|
=\frac{\Lambda}{6r_2^2}(r_2 - r_0)(r_2 - r_1)(r_3 - r_2), 
\end{equation} 
\begin{equation} 
\kappa_3 = \frac{1}{2}|V_{s}^\prime(r_3)|
=\frac{\Lambda}{6r_3^2}(r_3 - 
r_0)(r_3 - r_1)(r_3 - r_2). 
\end{equation}

The requirement of vanishing second fundamental form at $\Sigma$
minus the two conical singularities at the two horizons implies
that the 
transition can only occur at two sections of constant values of 
imaginary time $\tau$ glued at the two horizons. The 3-surface
$\Sigma$ has topology $S^2 \times S^1$.  To form a constrained
gravitational instanton, one can have two 
cuts at $\tau = consts.$ between $r = r_2$ and $r = r_3$. Then 
the $f_2$-fold cover turns 
the $(\tau - r)$ plane into a cone with a deficit angle 
$2\pi (1-f_2)$ at the black hole horizon. In a similar way one 
can have an $f_3$-fold cover 
at the cosmological horizon. Both $f_2$ and $f_3$ can take any 
pair of real numbers with the relation  
\begin{equation} 
f_2 \beta_2 = f_3 \beta_3, 
\end{equation} 
where $\beta_2 = 2\pi \kappa^{-1}_2$ and $\beta_3 = 2\pi
\kappa^{-1}_3$.
If $f_2$ or $f_3$ is different from $1$, then the cone at the 
black hole or cosmological horizon will have an extra 
contribution to the action of the instanton. After the transition
to 
Lorentzian spacetime, the conical singularities will only affect 
the 
real part of the phase of the wave function, i.e. the probability
of the creation of the black hole. 
 
Since the integral of $K$ with respect to the $3$-area in the 
boundary term of the action (2) is the area increase rate along 
its normal, then the extra contribution due to the conical 
singularities can be considered as the degenerate form shown 
below 
\begin{equation} 
\bar{I}_{2,deficit} = - \frac{1}{8 \pi}\cdot 4\pi r_2^2\cdot 2\pi
(1 - f_2), 
\end{equation} 
\begin{equation} 
\bar{I}_{3,deficit} = - \frac{1}{8 \pi}\cdot 4\pi r_3^2\cdot 2\pi
(1 - f_3). 
\end{equation} 

The action due to the volume is
\begin{equation}
\bar{I}_v = -\frac{f_2 \beta_2 \Lambda}{6} (r^3_3 - r^3_2) \pm
\frac{f_2 \beta_2 Q^2}{2}(r^{-1}_2 - r^{-1}_3),
\end{equation}
where $+$ is for the magnetic case and $-$ is for the electric
case. This term disappears for the neutral case.

In the neutral case, the boundary date on the 3-surface $\Sigma$
will be $h_{ij}$. In the magnetic case, the boundary date is
$h_{ij}$ and $A_i$. The vector potential in turn determines the
magnetic charge, since it can be obtained by the magnetic flux,
or the integral of the gauge field $F$ over the $S^2$ space
sector. It is more convenient to choose a gauge potential
\begin{equation}
A = Q(1- \cos \theta) d\phi
\end{equation} 
to evaluate the flux.

In the electric case, the boundary date is $h_{ij}$ and the
momentum $\omega$ [11], which is canonically conjugate to the
electric charge and defined by
\begin{equation}
\omega = \int A,
\end{equation}
where the integral is around the $S^1$ direction. The most
convenient choice of the gauge potential for the calculation is
\begin{equation}
A = -\frac{iQ}{r^2}\tau dr.
\end{equation}

The wave function for the equator is the exponential of half the
negative of the
action. For the neutral and magnetic cases, one obtains the wave
function $\Psi (h_{ij})$ and $\Psi (Q,h_{ij})$.
For the electric case, what one obtains this way is
$\Psi (\omega, h_{ij})$ instead of $\Psi (Q, h_{ij})$. One can
get the wave function $\Psi (Q, h_{ij})$ for
 a given electric charge through the Fourier
transformation [10][11]
\begin{equation}
\Psi (Q, h_{ij}) = \frac{1}{2\pi} \int^{\infty}_{-\infty} d
\omega e^{i\omega Q} \Psi
(\omega, h_{ij}).
\end{equation}
This Fourier transformation is equivalent to a multiplication of
an extra factor
\begin{equation}
\exp \left (\frac{- f_2\beta_2Q^2  ( r_2^{-1} -
r_3^{-1})}{2}\right )
\end{equation}
to the wave function.
This makes the probabilities for magnetic and electric cases
equal, and thus recovers the duality between the magnetic and
electric black holes [11].

Finally, using the relations (17) and (11)-(14), one obtains the
probability for 
a spherically symmetric black hole creation
\begin{equation}
P_{s} \approx \exp(\pi (r^2_2 + r^2_3)).
\end{equation}
This is the exponential of one quarter of the sum of the black
hole and cosmological horizon areas, or the total entropy of the
universe.

The most remarkable fact is that the result is
independent of our choice of $f_2$ or $f_3$. It means the
manifold has a stationary action, therefore it can be qualified
as a constrained gravitational instanton, and it can be
used for the
$WKB$ approximation to the wave function. The same phenomenon
will occur to the
Kerr-Newman case as one will see later.

For the cases of the nonsingular, charged
or neutral, spherically symmetric instantons and the associated
black hole creations [6][7][8][9][10][11], all these instantons
lead
to the creation of pairs of black holes. For these cases one can
avoid the conical singularities by choosing $f_2 = f_3 = 1$,
since the two surface gravities are identical. However, their
results are the special cases of our general formula (26),
recalling that the degenerate
horizon should be counted twice.

The wave function for the spherically symmetric black hole can
also be found [5].

When $m = 0$ and $Q = 0$, it is reduced to the de Sitter case 
\begin{equation}
P_{0} \approx \exp\left (\frac{3\pi}{\Lambda} \right )
\end{equation}
and when $Q=0$ and $r_2 = r_3$, it is reduced to the Nariai case 
\begin{equation}
P_{m_c} \approx \exp\left (\frac{2\pi}{\Lambda} \right ).
\end{equation}
The formula (26) interposes the above values for the two extreme
cases of neutral black holes.

The probability is a decreasing function with respect to
parameter $m$ and $|Q|$. So the de Sitter universe is the most
probable one for the Planckian era in quantum cosmology, as is
expected.

\vspace*{0,3in}
  
\large \bf IV. The Kerr-de Sitter black hole 
\vspace*{0.3in} 
 
\rm  
\normalsize

Now let us discuss the creation of a rotating black hole
in the de Sitter space background. The Lorentzian metric of the
black hole spacetime is [17]
\begin{equation}
ds^2 = \rho^2(\Delta^{-1}_r dr^2 + \Delta^{-1}_\theta d\theta^2)
+ \rho^{-2}
 \Xi^{-2}
\Delta_{\theta} \sin^2 \theta (adt - (r^2 + a^2) d\phi)^2 -
\rho^{-2} \Xi^{-2}\Delta_r  (dt - a \sin^2 \theta d \phi)^2,
\end{equation}
where
\begin{equation}
\rho^2 = r^2 + a^2 \cos^2 \theta,
\end{equation}
\begin{equation}
\Delta_r = (r^2 + a^2)(1 - \Lambda r^2 3^{-1}) - 2mr + Q^2 + P^2,
\end{equation}
\begin{equation}
\Delta_{\theta} = 1 + \Lambda a^2 3^{-1} \cos^2 \theta,
\end{equation}
\begin{equation}
\Xi = 1 + \Lambda a^2 3^{-1}
\end{equation}
and $m, a, Q$ and $P$ are constants, $m$ and $ma$ representing
mass and  angular momentum. $Q$ and $P$ are electric and
magnetic charges.

One can factorize $\Delta_r$ as follows
\begin{equation}
\Delta_r = -\frac{\Lambda}{3} (r - r_0)(r - r_1)(r - r_2)(r -
r_3),
\end{equation}
where the roots $r_0, r_1, r_2$ and $r_3$ are in ascending order,
$r_2$ and
$r_3$ are the black hole and cosmological horizons. The roots
satisfy the following relations:
\begin{equation}
\sum_i r_i = 0,
\end{equation}
\begin{equation}
\sum_{i>j} r_i r_j = - \frac{3}{\Lambda} + a^2,
\end{equation}
\begin{equation}
\sum_{i>j>k} r_ir_jr_k = - \frac{6m}{\Lambda},
\end{equation}
\begin{equation}
\prod_i r_i = - \frac{3(a^2 + Q^2 + P^2)}{\Lambda}.
\end{equation}

In this section we shall concentrate on the neutral case with $Q
= P = 0$. The Newman case with nonzero electric or magnetic
charge will be differed to the next section.

The probability of the Kerr black hole creation, at the $WKB$
level, is the
exponential of the negative half of its corresponding constrained
gravitational instanton.  The only instanton which can be used to
join the Lorentzian sector at the quantum transition is the
complex spacetime obtained from the Lorentzian metric by a
substitution  $ t \longrightarrow -i\tau$ only. However, for
convenience of calculation, we can let $a$ to be imaginary, and
then the complex metric becomes Euclidean.
After we get the probability for the imaginary $a$ value, then we
can analytically continue back to real $a$ to obtain the
required probability.

In order to form a constrained gravitational instanton, one can
do the similar cutting, folding and covering at both
the black hole and cosmological horizons with $f_2$ and $f_3$
satisfying relation (17) as in the nonrotating case. We shall
freely switch back and forth between the real and imaginary
values of $a$ in the following
calculation to facilitate our interpretation. 

For the Kerr case, the topology of 3-surface $\Sigma$ is $S^2
\times S^1$. Their horizon areas are
\begin{equation}
A_2 = 4\pi (r^2_2 + a^2)\Xi^{-1},
\end{equation}
\begin{equation}
A_3 = 4\pi (r^2_3 + a^2)\Xi^{-1}.
\end{equation}
The  black hole and cosmological surface gravities are
\begin{equation}
\kappa_2 =\frac{\Lambda (r_2 - r_0)(r_2 - r_1)(r_3 - r_2)}{6\Xi
(r^2_2 + a^2)},
\end{equation}
\begin{equation}
\kappa_3 = \frac{\Lambda (r_3 - r_0)(r_3 - r_1)(r_3 - r_2)}{6\Xi
(r^2_3 + a^2)}.
\end{equation}

The actions due to the conical singularities are
\begin{equation}
\bar{I}_{2, deficit} = - \frac{\pi (r^2_2 + a^2)(1 - f_2)}{\Xi},
\end{equation}
\begin{equation}
\bar{I}_{3, deficit} = - \frac{\pi (r^2_3 + a^2)(1 - f_3)}{\Xi}.
\end{equation}
 
The action due to the volume is
\begin{equation}
\bar{I}_v = - \frac{f_2\beta_2 \Lambda}{6\Xi^2} (r^3_3 - r^3_2 +
a^2(r_3 - r_2)),
\end{equation}
where $\beta_2$ is defined as before.

If one naively takes the exponential of the negative of half the
total action (after the analytic continuation by the replacement
of $b$ by $a$), then the wave function for the
creation moment of a black hole with parameter $m$ and $a$ will
not be obtained. The
physical reason is that what one can observe is only the angular
differentiation, or the relative rotation of the two horizons.
This situation is similar to the case of a Kerr black hole in the
asymptotically flat background. There one can only measure the
rotation of
the black hole horizon from the spatial infinity. To find the
wave function for the given mass and angular momentum one has to
make the Fourier transformation
\begin{equation}
\Psi(m, a, h_{ij}) = \frac{1}{2 \pi}\int^{\infty}_{-\infty}
d\delta e^{i\delta
J \Xi^{-2}} \Psi(m, \delta, h_{ij}),
\end{equation}
where $\delta$ is the relative rotation angle for the time period
$f_2\beta_2$, which is canonically conjugate to the angular
momentum $J \equiv ma$; and the factor $\Xi^{-2}$ is due to the
time rescaling.
The angle difference $\delta$ can be evaluated
\begin{equation}
\delta = \int_0^{f_2\beta_2/2} d\tau (\Omega_2 - \Omega_3),
\end{equation}
where the angular velocities at the two horizons are
\begin{equation}
\Omega_2 = \frac{a}{r^2_2 + a^2},
\end{equation}
and
\begin{equation}
\Omega_3 = \frac{a}{r^2_3 + a^2}.
\end{equation}

The Fourier transformation is equivalent to adding an extra term
into the action for the constrained instanton, and then the total
action becomes
\begin{equation}
\bar{I} = - \pi(r^2_2 + a^2)\Xi^{-1} - \pi(r^2_3 + a^2)\Xi^{-1}.
\end{equation}
It is crucial to note that the action is independent of
$\beta_2$, and therefore we obtain the constrained instanton.
The probability of the Kerr black hole creation is
\begin{equation}
P_k \approx \exp  (\pi(r^2_2 + a^2)\Xi^{-1} + \pi(r^2_3 +
a^2)\Xi^{-1}).
\end{equation}
It is the exponential of one quarter of the two horizon areas, or
the total entropy of the universe.

\vspace*{0.3in} 
 
\large \bf V. The Newman-de Sitter black hole
\vspace*{0.3in} 
\rm
\normalsize

Now let us turn to the charged black hole case. The vector
potential can be written as 
\begin{equation}
A =\frac{ Qr(dt - a\sin^2\theta d\phi) + P \cos \theta (a dt -
(r^2 + a^2) d\phi)}{\rho^2}.
\end{equation}

We shall not consider the dyonic case below.

One can closely follow the neutral rotating case for calculating
the action of the corresponding constrained gravitational
instanton. The only difference is to add one more term due to the
electromagnetic field to the action of volume. For the magnetic
case, it is
\begin{equation}
\frac{f_2\beta_2 P^2}{2\Xi^2} \left ( \frac{r_2}{r^2_2+ a^2} -
\frac{r_3}{r^2_3 + a^2} \right )
\end{equation}
and for the electric case, it is
\begin{equation}
-\frac{f_2\beta_2 Q^2}{2\Xi^2} \left ( \frac{r_2}{r^2_2+ a^2} -
\frac{r_3}{r^2_3 + a^2} \right )
\end{equation}

In the magnetic case the vector potential determines the magnetic
charge, which is the integral  over the $S^2$ space sector.
Putting
all these contributions together one can find
\begin{equation}
\bar{I} = - \pi(r^2_2 + a^2)\Xi^{-1} - \pi(r^2_3 + a^2)\Xi^{-1}
\end{equation}

and the probability of the creation of a magnetically charged
black hole is
\begin{equation}
P_n \approx \exp  (\pi(r^2_2 + a^2)\Xi^{-1} + \pi(r^2_3 +
a^2)\Xi^{-1}).
\end{equation}

In the electric case, one can only fix the integral
\begin{equation}
\omega = \int A,
\end{equation}
where the integral is around the $S^1$ direction.
So, what one obtains in this way is $\Psi(\omega, a, h_{ij})$. In
order to get the wave function $\Psi(Q, a, h_{ij})$ for a given
electric charge, we have to repeat the procedure like the
Reissner-Nordstr$\rm\ddot{o}$m case. The Fourier transformation
is equivalent to adding one more term to the action
\begin{equation} 
\frac{f_2\beta_2 Q^2}{\Xi^2}\left ( \frac{r_2}{r^2_2+ a^2} -
\frac{r_3}{r^2_3 + a^2} \right ).
\end{equation}

Then we obtain the same formula for the electrically charged
rotating black hole creation as that for the magnetic one,
\begin{equation}
P_n \approx \exp  (\pi(r^2_2 + a^2)\Xi^{-1} + \pi(r^2_3 +
a^2)\Xi^{-1}).
\end{equation}

It is easy to show that the probability is an exponentially
decreasing function of the mass parameter, charge magnitude and
angular momentum,
and the de Sitter spacetime is the most probable Lorentzian
evolution at the Planckian era.

\vspace*{0.3in}

\large \bf VI. Discussion
\vspace*{0.3in} 
\rm
\normalsize

The result of this paper has shown that the probability of
the black hole creation is the exponential of the total entropy
of the universe. The entropy is equal to one quarter of the sum
of
the black hole and cosmological horizon areas.

The probability is an exponentially decreasing function in terms 
of the mass parameter, charge magnitude and angular momentum.
Since this is only the confirmation of the conjecture, the
result is no surprise. The only surprise is the fact that
our result is independent of the choice of $f_2$ or $f_3$ for
the formation of the constrained gravitational instantons. 

To get a meaningful result, one has to be careful to identify
the meaning of the wave function; so for the rotating case and
electrically charged black holes, one has to introduce
Fourier transformations into the calculation; otherwise the
result
becomes meaningless. It is interesting to note that Nature would
give us a beautiful result if our request is reasonable.

In quantum field theory, the temperature associated with a black
hole is well defined. By using the reciprocal of the Hawking
temperature as the period of the imaginary time, one can avoid
the conical singularity at the horizon.
However, if we remain only at thermodynamics level, and if one
considers the reciprocal of the period for the constrained
gravitational instanton as an effective temperature, then from
the
calculation, it seems the temperature can be taken quite
arbitrarily. We appear to overcome the obstacle that the
temperature of the black hole and cosmological horizons, in
general, are different. This makes our calculation feasible.
Temperature is a very subtle concept even in special relativity,
let alone in general relativity. A thorough discussion about
temperature is beyond the scope of this paper. However, the
concept of entropy is very clear in any case.

Our calculation has also very clearly shown that the
gravitational entropy is associated with topology of spacetime,
as Hawking emphasized many times [18].

From the no-hair theorem, a stationary black hole in the de
Sitter
spacetime background can only have three parameters, mass, charge
and angular momentum, so the problem of the quantum creation of
a single black hole at the birth of the universe is completely
resolved. 

\rm  
\normalsize

\vspace*{0.1in}

\bf Acknowledgment:

\vspace*{0.1in} 
\rm 
 
I would like to thank Prof. G. Coyne of Specola Vaticana and
Prof. R. Ruffini of Rome University for their hospitality.  
\vspace*{0.1in}

\bf References: 
 
\vspace*{0.1in} 
\rm

1. S.W. Hawking, \it Mon. Not. R. Astro. Soc. \rm
\underline{152}, 75 (1971).

2. B. Carr and S.W. Hawking, \it Mon. Not. R. Astro. Soc. \rm
\underline{168}, 399 (1974).

3. Z.C. Wu, \it Phys. Rev. \bf D\rm\underline{30}, 286 (1984). 

4. L.Z. Fang and M. Li, \it Phys. Lett. \rm \bf B\rm
\underline{169}, 28 (1986).

5. Z.C. Wu, in \it Proceeding of the Fourth Marcel Grossman
Meeting \rm edited by R.Ruffini (North Holland, Amsterdam, 1986).

6. R. Bousso and S.W. Hawking, \it Phys. Rev. \rm \bf D\rm
\underline{52}, 5659 (1995).

7. P. Ginsparg and M.J. Perry,  \it Nucl. Phys. \rm \bf B\rm
\underline{222}, 245 (1983).

8. F. Mellor and I. Moss,   \it Phys. Lett. \rm \bf B\rm 
\underline{222}, 361 (1989); \it Class. Quantum Grav. \rm \bf
\underline{6}\rm, 1379 (1989). 

9. I.J. Romans, \it Nucl. Phys. \rm \bf B\rm \underline{383},
395 (1992).

10. R.B. Mann and S.F. Ross, \it Phys. Rev. \bf D\rm
\underline{52}, 2254
(1995).  

11. S.W. Hawking and S.F. Ross,  \it Phys. Rev. \bf D\rm 
\underline{52}, 5865 (1995).

12. J.B. Hartle and S.W. Hawking, \it Phys. Rev. \rm \bf D\rm
\underline{28}, 2960 (1983).

13. G. Hayward and J. Louko, \it Phys. Rev. \rm \bf D\rm 
\underline{42}, 4032 (1990). 

14. X.M. Hu and Z.C. Wu, \it Phys. Lett. \rm \bf B\rm 
\underline{149}, 87 (1984). 

15. G.W. Gibbons and J.B. Hartle, \it Phys. Rev. \rm \bf D\rm 
\underline{42}, 2458 (1990). 
 
16. S.W. Hawking, \it Nucl. Phys. \rm \bf B\rm \underline{239}, 
257 (1984).

17. G.W. Gibbons and S.W. Hawking, \it Phys. Rev. \bf D\rm 
\underline{15}, 2738 (1977). 

18. S.W. Hawking, in \it General Relativity: An Einstein
Centenary Survey, \rm eds. S.W. Hawking and W. Israel, (Cambridge
University Press, 1979).

\vspace*{0.6in}

\end{document}